\documentclass[final,3p,times,twocolumn,natbib]{elsarticle}
\usepackage{graphicx}
\usepackage{natbib}
\usepackage{hyperref}
\usepackage{amssymb}
\usepackage{lineno}
\usepackage{xcolor}

\journal{Physics Letter B}

\begin{document}

\begin{frontmatter}



\title{Competing asymmetric fusion-fission and  quasifission in neutron-deficient sub-lead nuclei}% in heavy-ion induced reactions


\author[a,b]{Shilpi~Gupta}
\author[a,b]{K. Mahata}
\ead{kmahata@barc.gov.in}

\author[a,b]{A.~Shrivastava}
\author[a]{K.~Ramachandran}

\author[a,b]{S.K.~Pandit}
\author[a,b]{P.C.~Rout}
\author[a]{V.V.~Parkar}

\author[b,c]{R.~Tripathi}
\author[a]{A.~Kumar}

\author[a,b]{B.K.~Nayak}
\author[a]{E.T.~Mirgule}
\author[a,b]{A.~Saxena}
\author[a]{S.~Kailas}

\author[d]{A.~Jhingan}
\author[e,f]{A.K. Nasirov}
\author[f]{G.A. Yuldasheva}
\author[g]{P. N. Nadtochy}
\author[h]{C. Schmitt}

\address[a]{Nuclear Physics Division, Bhabha Atomic Research Centre, Mumbai - 400085, India}
\address[b]{Homi Bhabha National Institute, Anushaktinagar, Mumbai - 400094, India}
\address[c]{Radiochemistry Division, Bhabha Atomic Research Centre, Mumbai - 400085, India}
\address[d]{Inter University Accelerator Centre, Aruna Asaf Ali Marg, New Delhi-110067, India}
\address[e]{Bogoliubov Laboratory of Theoretical Physics, JINR, Dubna, Russia}
\address[f]{Institute of Nuclear Physics, Tashkent, Uzbekistan}
\address[g]{Omsk State Technical University, Mira prospekt 11, 644050 Omsk, Russia}
\address[h]{Institut Pluridisciplinaire Hubert Curien (IPHC), CNRS/IN2P3, 23 rue du Loess, B.P. 28, F-67037 Strasbourg, France}

\date{\today}

\begin{abstract}
To disentangle the role of shell effects and dynamics, fission fragment mass distributions  of $^{191}$Au, a nucleus in the newly identified island of mass asymmetric fission in the sub-lead region, have been measured down to excitation energy of $\approx$20 MeV above the fission barrier via two different entrance channels, viz. $^{16}$O+$^{175}$Lu and $^{37}$Cl+$^{154}$Sm reactions. Apart from having signature of the shell effects in both the cases, clear experimental evidence of quasifission has been observed in the mass distributions of the Cl induced reaction, that has also been substantiated by the theoretical calculations.  This crucial evidence along with a systematic analysis of available experimental data has revealed that the dynamics in the entrance channel has significant influence on most of the reactions used earlier to explore the persistence of recently discovered mass asymmetry in $\beta$-delayed fission at low energy in this mass region, ignoring which might lead to ambiguity in interpreting the heavy-ion data.
\end{abstract}


\end{frontmatter}

%\linenumbers 

%\section*{Introduction}

Understanding nuclear fission, which represents a large scale collective phenomena known to be governed by the delicate interplay of the macroscopic (liquid drop)
aspects and the microscopic (shell) effects, continues to be challenging.  Unambiguous experimental information is crucial  for accurate modeling of the shell effects and the  dynamical aspects in fission. Reliable knowledge of fission is not only important for the fundamental research like nuclear physics and astrophysics, but also for the applications like nuclear energy and medicine. The richness and the complexity of the field along with the current status have been summarized  in the latest reviews~\cite{schmidt18,andreyev17,kailas14}.

Unexpected observations of mass-asymmetric fission in $^{180}$Hg~\cite{andreyev10} and multimodal fission in $^{194,196}$Po, $^{202}$Rn~\cite{ghys14}, populated just above the fission barrier in $\beta$-decay at ISOLDE-CERN, have given the opportunity to  test the knowledge gained in the actinide region.   
%By ascribing these observations to a relatively small microscopic effects that make the fission saddle point and the nearby valley  mass-asymmetric, the calculations~\cite{andreyev10,ichikawa18,ichikawa12} based on the state of the art five-dimensional (5D)  macroscopic-microscopic model~\cite{moller01} have predicted a new island of mass-asymmetric  fission in the sub-Pb region~\cite{moller15}. 
The calculations~\cite{andreyev10,ichikawa18,ichikawa12} based on the state of the art five-dimensional (5D)  macroscopic-microscopic model~\cite{moller01} ascribes these observations to a relatively small microscopic effects that make the fission saddle point and the nearby valley  mass-asymmetric. Consequently,  a new island of mass-asymmetric  fission in the sub-Pb region has been predicted~\cite{ichikawa18,moller15}.  However, improved scission point model calculations~\cite{panebianco12,andreev16} emphasize the importance of the deformation dependent shell effects in the final fragments to explain these observations. Fully self-consistent models~\cite{warda12,mcdonnell14} correlate these observations to the shell structure of prescission configurations. Recent microscopic mean-field calculations~\cite{scamps18,scamps19}  based on the Hartree-Fock approach with BCS pairing correlations advocate a universal mechanism, octupole correlations induced by deformed shell
gaps, for the observations of mass-asymmetric fission in the sub-lead and actinide region. 
Some of the theoretical models predict a strong persistence of these single particle effects even at higher excitation energies~\cite{moller15,mcdonnell14}.

Due to extremely challenging experimental conditions, $\beta$-delayed fission studies are limited.
Heavy-ion induced fusion-fission route has also been exploited to study the mass-asymmetric fission and its evolution with excitation energy in neutron deficient sub-lead nuclei, viz. $^{179}$Au~\cite{tripathi15}, $^ {180,190}$Hg~\cite{nishio15} and $^{182}$Hg~\cite{prasad15} using beams of $^{35}$Cl, $^{36}$Ar and $^{40}$Ca, respectively. The deviations in the measured mass distributions from single Gaussian shapes  at  excitation energy $\approx$ 25 MeV above the fission barrier  were associated to the observed mass asymmetry in $\beta$-delayed fission at very low excitation energy~\cite{andreyev10}. Recently, multimodal nature (competing symmetric and asymmetric  compound nuclear contributions) has been inferred in fission of $^{178}$Pt populated via $^{36}$Ar+$^{142}$Nd reaction~\cite{tsekhanovich19}.  
Heavy-ion induced reaction also provides the opportunity to study the possible link between the sub-Pb and the actinide region~\cite{gupta19}.

Use of heavy-ion beams  not only brings in higher excitation energy and angular momentum ($\ell$), it also opens the possibility of  quasifission, which might complicate the interpretation of the experimental observations substantially. The quasifission, a non-compound (non-equilibrated) nuclear process is being studied experimentally~\cite{itkis15,rietz13,banerjee19} as well as theoretically~\cite{aritomo04,zagrebaev05,nasirov11} with great vigor as it hinders formation of super-heavy elements. It strongly depends on the  entrance channel parameters like charge product (or mass asymmetry), deformation of the colliding nuclei, shell closure and neutron excess in addition to the compound nucleus (CN) fissility. On the lighter side of the explored map~\cite{rietz13}, evidence of quasifission has been found in $^{202}$Po (Z = 84), formed in $^{34}$S+$^{168}$Er reaction having  target projectile charge product ($Z_pZ_t$) as low as 1088~\cite{rafiei08}. 
%This aspect {\color{blue}has either been ignored or not considered in the analysis} in the  heavy-ion induced reactions having $Z_pZ_t$ in the range 1054 to 1200 used to study the presence of asymmetric fission in nuclei with Z$\leq$80~\cite{nishio15,prasad15,tripathi15,tsekhanovich19}. 
%However, the role of quasifission in asymmetric fission in the sub-lead region has not been investigated in the earlier studies using heavy-ion induced reactions having $Z_pZ_t$ in the range 1054 to 1200~\cite{nishio15,prasad15,tripathi15,tsekhanovich19}. 
Although the possible presence of quasifission was not ruled out   in  $^{40}$Ca+$^{142}$Nd reaction~\cite{prasad15}, its exact nature and extent in the sub-Pb region remained unexplored.
%~\cite{nishio15,tripathi15,tsekhanovich19}. 
Investigation of this aspect is essential for an accurate  modeling of the excitation energy dependence of  the microscopic effects.  Particularly, ignoring quasifission might lead to ambiguity in the inferred multimodal fission  in this region.%~\cite{tsekhanovich19}. 
So far, only  a few experimental data is available in the sub-Pb region and there are contradictory predictions from the theoretical  models. More measurements are required to verify the predicted generic nature of asymmetric fission~\cite{moller15} and to refine the theoretical models. 

In this Letter, we present measurements of fission fragment mass distributions of $^{191}$Au, populated using two different entrance channels $^{16}$O+$^{175}$Lu ($Z_pZ_t$ = 568) and $^{37}$Cl+$^{154}$Sm ($Z_pZ_t$ = 1054)  to understand the origin of mass-asymmetric fission in heavy-ion induced reactions in the sub-Pb region, by disentangling the role of the shell effects and dynamics in the entrance channel.

%\section{Experimental details}\label{expt}
Pulsed beams of $^{16}$O and $^{37}$Cl from the BARC-TIFR Pelletron-Linac Facility, Mumbai were bombarded on a 280 $\mu$g/cm$^2$ thick $^{175}$Lu (97.41\% enriched) target on a 150 $\mu$g/cm$^2$ thick Al backing and a 200 $\mu$g/cm$^2$ thick $^{154}$Sm ($>$ 99\% enriched) target on a 550 $\mu$g/cm$^2$ thick Al backing, respectively. 
Fission fragments time-of-flights (TOF) with respect to the arrival of the beam pulse, positions (x,y)  and energy losses were recorded using two large area (12.5 $\times$ 7.5 cm$^2$) position sensitive multiwire proportional counters (MWPCs)~\cite{jhingan15} kept at a distance of 24 cm from the target, covering an angular range of 30$^{\circ}$ each. 
 To detect both the fragments in coincidence, the detectors were placed around the beam axis at $\theta_1$ = -50$^{\circ}$, $\theta_2$ = 107$^{\circ}$ for $^{16}$O+$^{175}$Lu with target facing the beam and at $\theta$ = $\pm$64$^{\circ}$ for $^{37}$Cl induced reaction with backing facing the beam.

\begin{figure}
\includegraphics[trim = 0mm 0mm 0mm 0mm, clip]{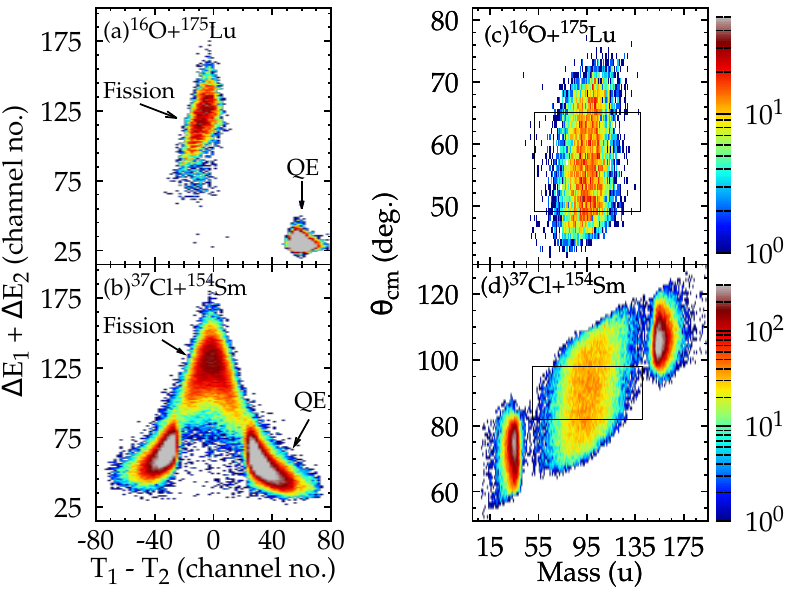}%,width=0.4\textwidth
\caption{\label{fig:mad} Time of flight difference (${\rm T}_{1}-{\rm T}_{2}$) vs energy loss ($\Delta{\rm E_1}+\Delta{\rm E_2}$) spectra  used to separate fission from quasi-elastic (QE) events for (a)$^{16}$O+$^{175}$Lu at E$_{\rm lab.}$=82.8 MeV and (b)$^{37}$Cl+$^{154}$Sm  reaction at E$_{\rm lab.}$=166.4 MeV. The corresponding mass angle distributions along with the angular cut (rectangular box) used to obtain the mass distributions are shown in  (c)  and (d), respectively.}
\vspace{-0.5cm}

\end{figure}

\begin{figure}
\includegraphics[trim = 0mm 0mm 0mm 0mm, clip]{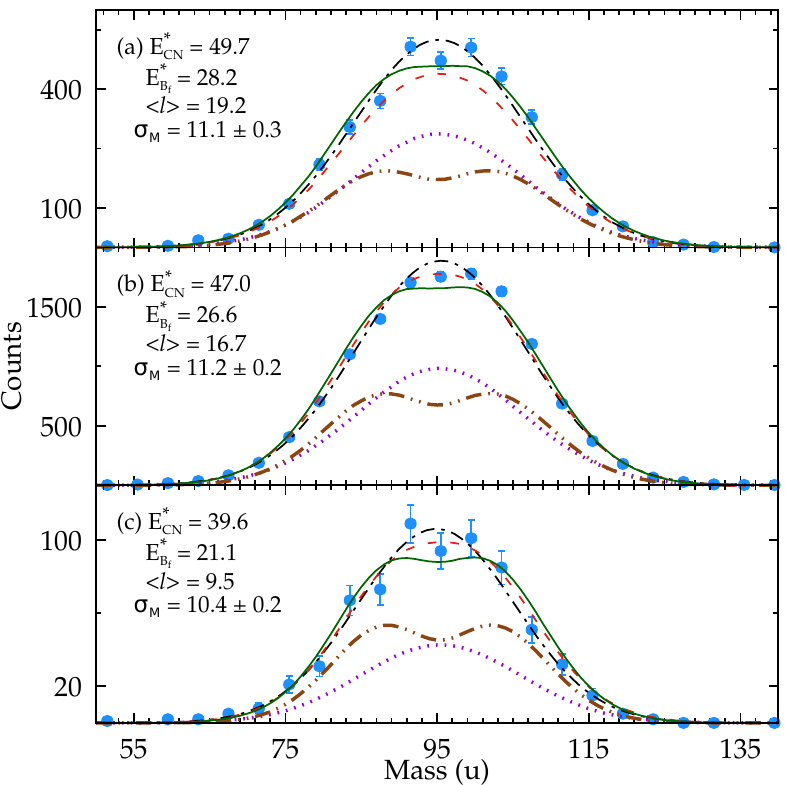} %,width=0.45\textwidth
\caption{\label{fig:OLu} The experimental fission fragment mass distributions (blue filled circles) for $^{16}$O+$^{175}$Lu reaction at different excitation energies are compared with the predictions of total (green continuous line) along with the symmetric (purple dotted) and asymmetric (brown dot-dot-dash) components of GEF code~\cite{schmidt16}. The sum of  25\% asymmetric and 75\% symmetric components are shown in red dashed line. The black dash-dotted lines are the single Gaussian fits. The excitation energy of the compound nucleus (E$^*_{\rm CN}$) and the effective excitation energy above the fission barrier (E$^*_{\rm B_f}$) (see text) in MeV are noted along with the estimated average angular momentum ($\langle\ell\rangle\hbar$)  and width ($\sigma_{\rm M}$) of the single Gaussian fits.}
\vspace{-0.5cm}

\end{figure}

\begin{figure}
\includegraphics[trim = 0mm 0mm 0mm 0mm, clip]{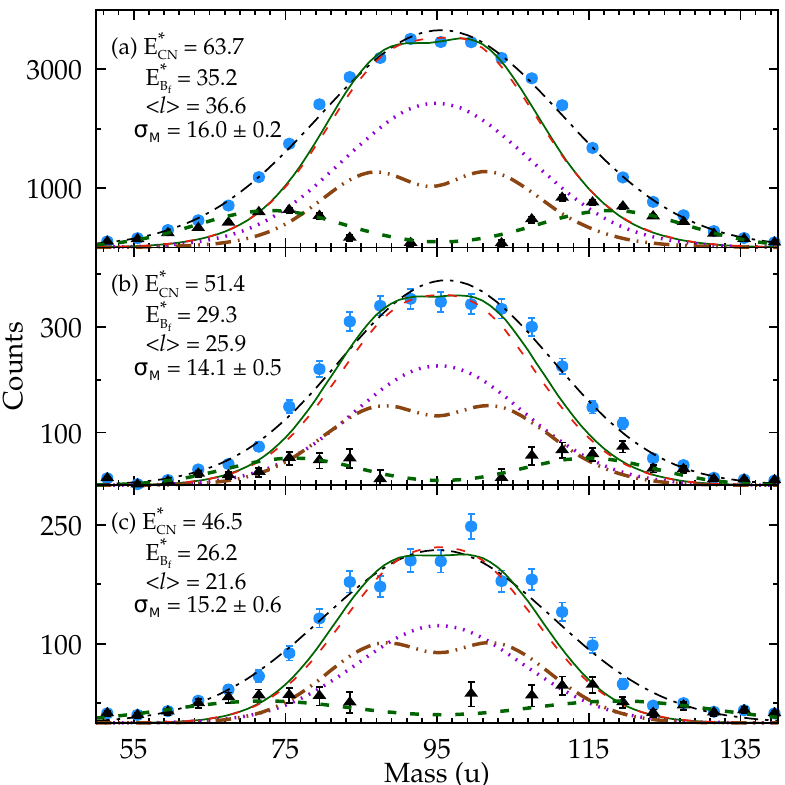}
\caption{\label{fig:ClSm} Same as Fig.~\ref{fig:OLu} except for $^{37}$Cl+$^{154}$Sm reaction. The differences between the measured distributions and the GEF predictions are also shown as filled triangle along with sum of two Gaussian fits (green dashed lines).}
\vspace{-0.5cm}

\end{figure}

%\section{Data Analysis}\label{result}
The detected fragment velocity vectors were calculated from the TOF and position information. The fission events were selected by putting two dimensional gates in the TOF difference vs energy loss spectra shown in Fig.~\ref{fig:mad} (a-b).  The correlations between the folding and azimuthal angles as well as between parallel and perpendicular components of the velocity onto the beam axis for the selected fission events confirm the absence of transfer induced (incomplete momentum transfer) events.
Fragment mass distributions were deduced using the TOF difference method~\cite{choudhury99}. The mass resolution ($\sigma$) was estimated from the elastic peak to be 2.8 u. Small corrections in the fragment mass due to their energy loss in the target and backing were obtained on an event-by-event basis in an iterative manner, taking the energy loss information from SRIM~\cite{ziegler10} for all the possible fragments. Typical correction in the width due to energy loss are about 4.5\%  and 2\%  for  $^{16}$O+$^{175}$Lu and $^{37}$Cl+$^{154}$Sm systems, respectively. 
Typical mass-angle correlation plots  are shown in Fig.~\ref{fig:mad} (c-d). No significant mass angle correlation has been observed for both the systems at all energies studied. Mass angle correlation is also not expected as the fissility parameters of the present systems are well below the experimentally determined threshold only above  which mass angle correlation is observed~\cite{rietz13}. The experimental mass distributions (Fig.~\ref{fig:OLu}~and~\ref{fig:ClSm}) were obtained by projecting the mass angle correlations with angular cut (see Fig.~\ref{fig:mad} (c-d)) to remove the bias due to geometrical acceptance of the detection setup. 
%The distributions of the complementary fragment are found to be similar and added to the above distribution to improve the statistics.

%\section{Result}\label{result}

For a purely macroscopic potential energy surface, the fragment mass distribution of CN fission is expected to be a Gaussian in shape. Even though the overall mass distribution could be fitted well with single Gaussians, deviations are observed at the middle of the distribution in all cases (see  Fig.~\ref{fig:OLu}~and~\ref{fig:ClSm}).  
The experimental mass distributions are compared with the predictions of the semiempirical model GEneral description of Fission observables (GEF)~\cite{schmidt16}  with global parameter values. This model is used to describe the observables of spontaneous fission  as well as CN fission for a given excitation energy (E$_{\rm CN}^*$) and average angular momentum ($\langle\ell\rangle $). The $\langle\ell\rangle $ values were calculated using the coupled channels code CCFULL~\cite{hagino99}.
The fusion excitation functions for the present system is not available. The data for similar system, $^{16}$O+$^{176}$Yb~\cite{rajbongshi16}, was fitted to constrain the potential parameters for the CCFULL calculations. As can be seen from Fig.~\ref{fig:OLu}, there is a good agreement between the measured mass distributions and the model predictions for the $^{16}$O+$^{175}$Lu system. Particularly, the observed deviation from a  Gaussian shape at the middle of the distribution is also reproduced well by the model, in which microscopic corrections are already incorporated empirically. The GEF predicts 60\%, 49\% and 45\% of asymmetric compound nuclear contributions for E$^*_{\rm CN}$ = 39.6, 47.0 and 49.7 MeV, respectively. The experimental data is found to be less sensitive to the relative weight of the asymmetric to symmetric  component.  This might be due to the similar  overall widths of the predicted symmetric and asymmetric components. Use of 25\% asymmetric and 75\% symmetric contributions, as shown in Fig.~\ref{fig:OLu}, results in the best fits by reducing  the $\chi^2$ by only a factor of 2 as compared to the GEF predicted percentages.

Apart from showing similar deviations from Gaussian shapes at the middle, the mass distributions for the more symmetric system ($^{37}$Cl+$^{154}$Sm:  Fig.~\ref{fig:ClSm}) are found to be  broader than those for the asymmetric combination ($^{16}$O+$^{175}$Lu: Fig.~\ref{fig:OLu}). 
This could be due to larger angular momentum involved in the case of  heavier projectile as well as due to the presence of quasifission component. 
The estimated $\langle\ell\rangle$ values (see Fig.~\ref{fig:ClSm}), using CCFULL with potential parameters constrained by fitting the fusion excitation function for $^{40}$Ar+$^{154}$Sm reaction~\cite{reisdorf85}, are about 6$\hbar$ higher as compared to those for $^{16}$O+$^{175}$Lu system at similar E$^*_{\rm CN}$. For $^{16}$O+$^{175}$Lu system, with a variation of 10 MeV in E$^*_{\rm CN}$ and 10 $\hbar$ in $\langle\ell\rangle$, there is only a 6.5\% change in the measured mass width. This rules out a significant role of $\ell$ in increasing the width for $^{37}$Cl+$^{154}$Sm as compared to  $^{16}$O+$^{175}$Lu system at similar E$^*_{\rm CN}$ and reveals the presence of quasifission in the former case.
Though the shape of the distributions at the middle are well reproduced, the measured mass distributions are found to be much broader than the distributions predicted by the GEF (see Fig.~~\ref{fig:ClSm}), confirming the presence of quasifission. 
The estimated quasifission contributions, differences between the measured distributions and the GEF predictions, are found to  overlap significantly with the compound nuclear contributions.
The quasifission contribution is about 20\% of the total counts at all three energies. 
%\sout{The sum of three Gaussian fits to the measured distributions  (shown only for E$^*_{\rm  CN}$ = 51.4 MeV in Fig.~\ref{doubleG} (a)) with the estimated width for the CN contributions from Eq.~1  (discussed later)  also yields similar contributions for quasifission.}

\begin{figure}
\includegraphics[scale=1]{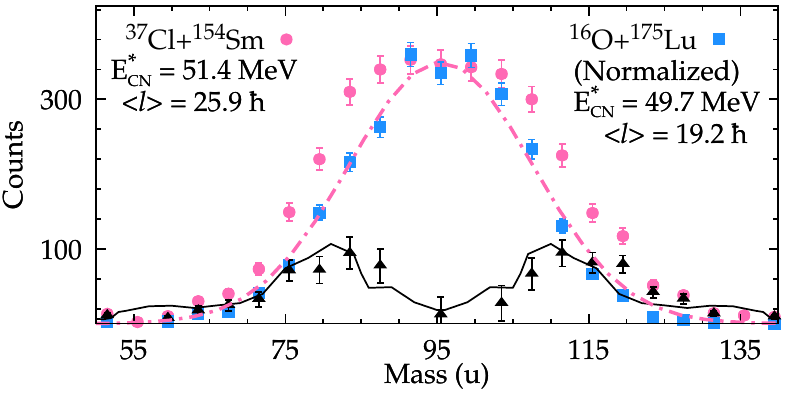}
\caption{\label{doubleG} The  difference (filled triangles) between the measured mass distributions for the two reactions (filled circles and squares) at similar ${\rm E^*_{CN}}$ and $\langle \ell \rangle$ is compared with the result of the dinuclear system (DNS) model calculation (continuous line) for quasifission in $^{37}$Cl+ $^{154}$Sm system. The  dot-dashed line is the expected distribution from the statistical relation (Eq. 1) for the $^{37}$Cl+ $^{154}$Sm system. %Sum of two Gaussian functions,  assuming only a single asymmetric CN fission mode, fits to the experimental mass distributions are shown in (b) and (c) as continuous lines.  The peak positions of the asymmetric components (dashed lines) are marked by dotted vertical lines.
}
\vspace{-0.5cm}

\end{figure}

%In the present measurement, it has been possible to match the  E$^*$ and $\langle\ell\rangle$, roughly. 
The mass distributions were also calculated for both the systems at similar  E$^*_{\rm  CN}$ (for the data shown in Fig.~2(a) and 3(b))  using the 4D Langevin dynamical model of CN evolution~\cite[and references therein]{nadtochy14,mazurek17}, taking the CN spin distributions from CCFULL. The one-body dissipation mechanism with the reduction coefficient $k_s$, obtained from the chaos theory~\cite{pal96} as well as $k_s$ = 1, were used to describe dissipation of the collective energy. The finite-range liquid drop model~\cite{sierk86} was used to calculate the potential energy. 
The calculated distributions do not show any significant difference between the two systems with similar E$^*_{\rm  CN}$.
Similar observations were made from the GEF calculations and  the statistical relation (Eq.~1; discussed later) as well. Hence, the difference between the two measured distributions (shown in Fig.~\ref{doubleG}) can be considered as the quasifission contribution. 
%As the estimated sticking time for the observed quasifission peak ($\sim$78~u), using the systematics for mass drift~\cite{rietz11,shen87,toke85}, and the half rotation time are similar, the mass-angle correlation is expected to be absent. 

To get a deeper insight, the distribution of the quasifission products were calculated in the framework of the dinuclear  system model~\cite{nasirov05,kim15} by solving the transport master-equation with the transition coefficients  which depend on the single-particle energies and occupation numbers of the interacting nuclei  (see Ref. \cite{thakur17}). The transition coefficients are sensitive to the shape and orientation of the interacting nuclei and $\ell$ distribution. The change of the excitation energy of the dinuclear system due to the  change of the intrinsic energy of its interacting fragments at the proton and  neutron transfer is taken into account. The DNS model predictions of 22\% qasifission for $^{37}$Cl+ $^{154}$Sm reaction and negligibly small quasifission contribution for $^{16}$O+$^{175}$Lu reaction are in good agreement with the experimental observations. 
%The experimentally observed quasifission contributions ($\sim$20\%) for the $^{37}$Cl+ $^{154}$Sm reaction  could be reproduced with a lifetime of  the dinuclear system (t$_{DNS}$) of 20$\times 10^{-22}$s. It results negligibly small quasifission contribution for $^{16}$O+$^{175}$Lu system.   
The calculated distribution of the quasifission products for the  $^{37}$Cl+ $^{154}$Sm (E$^*_{\rm  CN}$ = 51.4 MeV) reaction  is also in good agreement with the experimentally obtained distribution as shown in Fig.~\ref{doubleG}. Shell effects in the emerging  light fragments ($Z$=32--34 and $N$= 46--48) of the dinuclear system found to persist  at these energies and influence the outcome.

\begin{figure}
\includegraphics[trim = 0mm 0mm 0mm 0mm, clip]{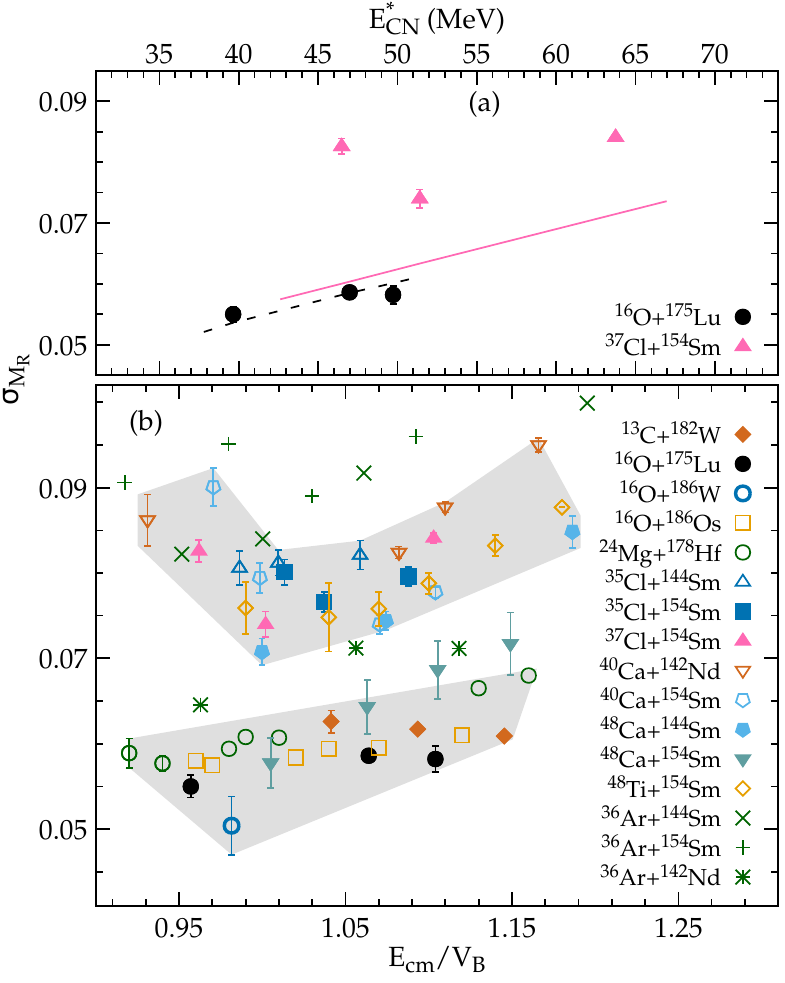}
\caption{\label{width} Experimental mass widths relative to CN mass ($\sigma_{\rm MR}$) for (a) $^{191}$Au in $^{16}$O+$^{175}$Lu and $^{37}$Cl+$^{154}$Sm reactions and (b) near by nuclei in heavy-ion induced reactions~\cite{nishio15,prasad15,tripathi15,tsekhanovich19,rafiei08,knyazheva07}. The dashed line is the  fit by the Eq.~\ref{sigma} to the data for $^{16}$O+$^{175}$Lu system assuming  compound nucleus fission only and the solid line is the estimated widths for $^{37}$Cl+$^{154}$Sm system using the same parameters.  The region of C,O,Mg and Cl,Ca (except $^{48}$Ca+$^{154}$Sm, see text) are shaded separately to highlight the difference among them in (b).}
\vspace{-0.6cm}
\end{figure}

Since the deviations from single Gaussians are small, we have also examined the widths of the fitted Gaussian to study the role of the entrance channel dynamics. 
The ratio ($\sigma_{\rm MR}$) of widths  of the fitted Gaussians ($\sigma_{\rm M}$) to the CN mass (A$_{\rm CN}$) are  plotted in Fig.~\ref{width} as a function of ${\rm E^*_{CN}}$ and  ${\rm E_{cm}/V_B}$, where ${\rm E_{cm}}$ and ${\rm V_B}$ are the energy in the centre of mass and the Coulomb barrier, respectively.  While the experimental mass widths for $^{16}$O+$^{175}$Lu system is found to increase monotonically with increasing energy, the mass width shows a  increase with decreasing energy  below the  Coulomb barrier for $^{37}$Cl+$^{154}$Sm system,  characteristic to quasifission involving deformed targets~\cite{lin12}. The mass widths  are also found to be larger for $^{37}$Cl+$^{154}$Sm system as compared to those for $^{16}$O+$^{175}$Lu system. 

For macroscopic potential energy surface, width of the fragment mass distribution (${\sigma_{\rm MR}}$) in CN fission can be statistically described as~\cite{itkis98},
\begin{equation}
\label{sigma}
{\sigma_{\rm MR}^2}={\lambda}{\textrm{T}}+{\kappa}{\langle \ell^{2} \rangle}.
\end{equation}
The temperature at the saddle point (T) is defined as
${\textrm T}=\sqrt{{\textrm E^*_{\rm B_f}}/a}.$
The average excitation energy at the saddle point is given as $\textrm E^*_{\rm B_f} = \textrm E_{\rm {CN}}^*-\textrm B_f({\langle \ell \rangle})-\textrm E_{pre}-\textrm E_{rot},$
where E$_{\rm CN}^*$, B$_f(\langle\ell\rangle)$, E$_{pre}$ and E$_{rot}$ are CN excitation energy, fission barrier at $\langle\ell\rangle$, average 
energy removed by the pre-saddle  neutrons and rotational energy of the CN, respectively. The value of the level density parameter ($a$) is taken as   A$_{\rm CN}/9$. The rotating finite range  model (RFRM)~\cite{sierk86} has been used to calculate E$_{rot}$ and the change in the predicted fission barrier~\cite{mollerb15} due to $\ell$. The E$_{pre}$ values are estimated using the statistical model code PACE~\cite{gavron80,mahata15}.

Assuming that the statistical description is valid for the more asymmetric system, the experimental widths for the  $^{16}$O+$^{175}$Lu system are fitted  to obtain the coefficients of the above expression. The mean square values of angular momentum (${\langle \ell^2 \rangle}$) are obtained from CCFULL calculation as discussed earlier. The T and  ${\langle \ell^2 \rangle}$ range of the present measurement are not sufficient to constrain both the coefficients simultaneously.  The value of ${\kappa}$  was kept same ((1.23${\pm}$0.24)${\times}$10$^{-6}$) as used for the near by system $^{16}$O+$^{186}$W~\cite{knyazheva07}. The best fit could be obtained with ${\lambda}$=(2.77${\pm}$0.08)${\times}$10$^{-3}$.  The value of ${\lambda}$ and ${\kappa}$ are in good agreement with the systematics~\cite{adeev88}. As can be seen in Fig.~\ref{width}, the calculated values of ${\sigma_{\rm MR}}$  using the same coefficients  for $^{37}$Cl+$^{154}$Sm system are  much smaller than the experimentally obtained widths. The observed mass widths can not be reproduced by reasonable variation of the parameters and estimated  ${\langle \ell^2 \rangle}$. This observation further confirms the significant presence of quasifission.

We have compared the experimental mass widths of neutron deficient nuclei near Pb~\cite{nishio15,prasad15,tripathi15,tsekhanovich19,rafiei08,knyazheva07} in Fig.~\ref{width}~(b).  The fitted mass widths for most of the heavier projectile ($^{35,37}$Cl,$^{40,48}$Ca and $^{48}$Ti) induced and lighter projectile ($^{13}$C, $^{16}$O and $^{24}$Mg) induced reactions show distinctly different behavior as shown by the shaded regions. In general, Cl, Ca and Ti induced reactions involving both spherical as well as deformed targets exhibit significantly larger widths as compared to C - Mg induced reactions. Further, all the systems involving $^{154}$Sm (deformed) target with heavy beams show an increase in the width with decreasing energy below the Coulomb barrier. 
%However, in spite of having mass angle correlation~\cite{knyazheva07}, the mass widths for $^{48}$Ca+$^{154}$Sm system are found to be  smaller compared to those for other heavier  ion induced reactions and are comparable to those of lighter ion induced reactions. 
In case of neutron rich $^{48}$Ca+$^{154}$Sm system~\cite{knyazheva07}, the quasifission exhibits signature of fast time scale, i.e., observation of mass-angle correlation in asymmetric splits, which are clearly separated from the fusion-fission (symmetric) products. The widths of the symmetric distributions are found to be comparable to those of lighter ion induced reactions, thus having no significant contribution from quasifission in the symmetric region. While no such distinctly separate quasifission contribution is observed for $^{48}$Ca+$^{144}$Sm and $^{40}$Ca+$^{154}$Sm~\cite{knyazheva07}, widths of the symmetric distributions for these systems are found to be larger as compared to those for $^{48}$Ca+$^{154}$Sm system and other lighter ion induced reactions, indicating the presence of slow quasifission in these neutron deficient combinations. This also suggests a strong role of N/Z on the nature of quasifission.
In case of $^{36}$Ar+$^{142}$Nd,$^{144,154}$Sm~\cite{tsekhanovich19,nishio15}, the measured mass distributions shows large deviation from a single Gaussian distribution hence we have plotted the square root of the variance. While the data for  $^{36}$Ar+$^{142}$Nd are found to lie below the shaded region for heavier projectiles and are in agreement with GEF prediction~\cite{gupta19}, the data for $^{36}$Ar+$^{144,154}$Sm are found to be much higher. The above comparison indicates that most of the systems involving heavier projectile are having contribution from the quasi-fission process.    

In summary, the fragment mass distribution in fission of $^{191}$Au, formed via two different entrance channels have been measured down to excitation energy of $\approx$20 MeV above the fission barrier.  Observed deviations from the Gaussian shape around symmetric mass split in both the cases at all energies indicate the presence of shell effect. However, the experimental data suggest that the shell effect or its persistence with excitation energy is much weaker than the theoretical predictions~\cite{moller15,mcdonnell14}. 
The experimental mass distributions for $^{37}$Cl+$^{154}$Sm system are found to be much broader than those for $^{16}$O+$^{175}$Lu system at similar E$^*_{\rm  CN}$ and $\langle \ell \rangle$. 
Such a difference is not expected  in the decay of compound nucleus, according to the  statistical relation (Eq.~\ref{sigma})~\cite{itkis98}, semi-emprical code GEF~\cite{schmidt16} as well as the 4D Langevin dynamical model~\cite{nadtochy14,mazurek17}. The mass width  for $^{37}$Cl+$^{154}$Sm system was found to increase with decreasing energy below the Coulomb barrier. These results provide conclusive evidence of substantial presence of quasifission for the more symmetric entrance channel. The quasifission contribution is found to overlap with the compound nuclear contributions. This makes the inference of asymmetric and multimodal fission ambiguous in reactions involving projectiles with Z$\geq$17.  It is also evident from the systematic analysis of  the available experimental data that there is a significant presence of quasifission in the reactions involving heavier projectiles (Z$\geq$17) with spherical as well as deformed targets used to investigate fission of neutron deficient sub-Pb nuclei. Such a substaintial presence of quasifission was not anticipated in earlier studies~\cite{nishio15,prasad15,tripathi15,tsekhanovich19}. The Dinuclear system (DNS) model calculation, which reproduces the observed  quasifission probability and its distribution, has revealed the persistence of shell effects in the emerging  light fragments of the dinuclear system. Present study demonstrates for the first time that not only the shell effects, but the dynamics in the entrance channel also has a significant
role in influencing the fission of nuclei in the newly identified island of mass asymmetry. Both these aspects needs to be considered to interpret heavy-ion data unambiguously.
Present observations could provide  crucial inputs to the advanced theoretical models being developed to understand the influence of the shell effects and dynamics in fission.
%Present observations provide  crucial inputs for accurate theoretical modeling of the shell effects and dynamics in fission.

%\section*{Acknowledgments}
Authors thank the staff of BARC-TIFR Pelletron-Linac Facility for smooth operation. One of authors (A.K.N) is grateful to the DST-RFBR (Project 17-52-45037) for the partial support. %Authors also thank Dr. C. Schmitt for careful reading of the manuscript and useful suggestions. 

%\bibliography{mybib}

\end{document}